\documentclass{article}

\usepackage[margin=1in]{geometry}
\usepackage{natbib}
\usepackage{eucal}
\usepackage{amsmath, amssymb, amsthm}
\usepackage{graphicx}
\usepackage{float}
\usepackage{subcaption} 

\let \mc   = \mathcal
\let \tu   = \textup
\let \eq   = \equiv

\let \sube = \subseteq

\let \ta   = \rightarrow
\let \at   = \leftarrow

\newcommand{\pa}[2]{\ensuremath{\tu{pa}_{#1}(#2)}}
\newcommand{\ch}[2]{\ensuremath{\tu{ch}_{#1}(#2)}}

\newcommand{\s}[1][0.5]{\ensuremath{\mkern #1 mu}}
\newcommand{\ind}{{\;\perp\!\!\!\perp\;}}
\newcommand{\istate}[4]{\ensuremath{#1 \ind #2 \mid #3 \; [\, #4 \,]}}

\title{An extensive simulation study evaluating the interaction of resampling techniques across multiple causal discovery contexts}

\author{Ritwick Banerjee \and Bryan Andrews \and Erich Kummerfeld}
\date{\today}

\begin{document}

\maketitle

\begin{abstract}
Despite the accelerating presence of exploratory causal analysis in modern science and medicine, the available non-experimental methods for validating causal models are not well characterized. One of the most popular methods is to evaluate the stability of model features after resampling the data, similar to resampling methods for estimating confidence intervals in statistics. Many aspects of this approach have received little to no attention, however, such as whether the choice of resampling method should depend on the sample size, algorithms being used, or algorithm tuning parameters. We present theoretical results proving that certain resampling methods closely emulate the assignment of specific values to algorithm tuning parameters. We also report the results of extensive simulation experiments, which verify the theoretical result and provide substantial data to aid researchers in further characterizing resampling in the context of causal discovery analysis. Together, the theoretical work and simulation results provide specific guidance on how resampling methods and tuning parameters should be selected in practice.

\end{abstract}

\section{Introduction}
\label{sec:intro}

The growing prominence of causal analysis across scientific disciplines—including healthcare, policy development, social sciences, and beyond—has created significant methodological challenges. Researchers frequently face scenarios where conducting controlled experiments to isolate specific causal effects among numerous interacting variables is impractical or ethically unfeasible. Causal discovery methods have emerged as a potential solution, offering techniques to infer causal relationships from observational data. However, validating these inferred relationships remains a persistent challenge that has hindered broader acceptance of these approaches.

Unlike conventional machine learning and statistical modeling paradigms, which benefit from established evaluation frameworks (such as ROC-AUC scores and F1-scores for classification tasks, or $R^2$ and goodness-of-fit tests for statistical models), causal discovery lacks comparable standardized validation mechanisms. This methodological gap has contributed to skepticism regarding the reliability of causally discovered relationships and limited their integration into mainstream scientific practice.

Resampling techniques represent a promising approach for validating causal discovery algorithms. By applying causal methods to different subsets or variations of an observational dataset, researchers can compare the consistency of identified dependencies across resamples. This approach provides insights into the stability, reliability, and calibration of discovered causal structures. Despite their potential utility, however, the systematic application of resampling techniques for evaluating causal discovery algorithms remains understudied.

In this paper, we investigate best practices for applying resampling methods to evaluate the trustworthiness of causal discovery results. Through in silico experimentation, we generate synthetic data from known causal structures and systematically apply various causal discovery algorithms in conjunction with different resampling techniques. Additionally, we also provide formal theory that supports some of these findings. This controlled experimental framework allows us to assess the effectiveness of different evaluation approaches against ground truth, thereby establishing methodological guidelines for the validation of causal discovery results.

Our experimental results demonstrate that resampling techniques provide substantial benefits for graphical model structure learning across multiple dimensions. We find that ensembles created through bootstrapping and subsampling approaches improve precision and produce adequately calibrated models. In particular, ensemble methods based off of resampling techniques protect against adding erroneous edges. Notably, our analysis reveals interesting interplay between resampling strategies and tuning parameters, with different combinations yielding comparable performance outcomes. These benefits remain consistent across various network structures and densities, suggesting that resampling approaches offer a generalizable framework for enhancing graphical model estimation, particularly in high-dimensional settings with limited sample sizes.

\subsection{Related Work}

Only a few previous publications have investigated resampling and model stability in the context of discovering causal graphical models, most notably, \cite{jabbari2017obtaining} examined the calibration of directed edges in a bootstrapped version of the Really Fast Causal Inference algorithm and found that they were generally well calibrated. Further, \cite{wang2023confidence}
developed a framework to quantify uncertainty in causal discovery by constructing confidence sets for causal orderings in directed acyclic graphs. Additionally, \cite{debeire2024bootstrapaggregationconfidencemeasures} enhanced causal discovery for time series by introducing a bootstrap-based uncertainty estimation and bagging approach, demonstrating that it improves reliability and accuracy. Finally, \cite{kummerfeld2019simulationsevaluatingresamplingmethods} explored bootstrapping and $90\%$ subsampling as resampling techniques and how they affect the fGES algorithm and consequently calibration associated with fGES and the resampling techniques. This paper reports the results of a series of simulations conducted as an extension of the work by \cite{kummerfeld2019simulationsevaluatingresamplingmethods} by exploring a larger variety of resampling techniques, graph types, algorithms, and model tuning parameters.

\subsection{Key Contributions:}
\begin{itemize}
    \item We conducted and present results from a simulation study of resampling methods in the context of causal discovery, which is far more extensive than any previous simulation study of this topic.
    \item This study reveals via simulations that several resampling methods are approximately the same, namely 50\% subsample, and 100\% resampling but using the Effective Sample Size for calculations.
    \item The simulations demonstrate via both theory and simulations that bootstrapping without adjusting for effective sample size is approximately equivalent to adjusting for effective sample size and reducing the penalty discount by one half.
    \item We demonstrate with simulations that none of our tested penalty discounts or resampling approaches were universally superior to all other methods.
    \item The study demonstrates 90\% subsampling with penalty discount 2, and Bootstrapping with effective sample size adjustment with penalty discount 1, both performed well in many situations.
    \item Bootstrapping without effective sample size had particularly poor performance in almost all situations for the Best Order Score Search algorithm.
    \item Our simulations show that the complexity penalty of BIC is not automatically optimized for all sample sizes. In other words, users should consider modifying the weight of the penalty term (i.e. change the penalty discount) based on the empirical sample size (as well as the resampling method being employed, if any). 
\end{itemize}

\subsection{Outline}

This paper is organized as follows. Section 2 establishes the theoretical foundations necessary for our experiment, providing concise overviews of causal discovery methodologies, resampling techniques, and directed acyclic graphs (DAGs). We also introduce the key frameworks employed in our analysis, including Bayesian Information Criterion, likelihood ratio tests, and performance metrics such as Brier scores, Expected Calibration Error (ECE), precision, recall, and F1-score.

Section 3 presents our comprehensive analysis of resampling techniques for causal discovery validation. We detail our simulation design, which elaborates various causal structures used and the data generation processes, followed by a thorough examination of the experimental results. Section 4 contextualizes our findings within the broader landscape of causal discovery methodologies, discussing implications for researchers. Finally, Section 5 outlines promising directions for future work, identifying opportunities to extend our methodological framework and address challenges in causal discovery validation.

\section{Background}

    This section introduces relevant background and notation.

\subsection{Resampling}
    Resampling is the process of creating a new set of samples from an existing set of samples. There are a few common procedures used for resampling. In this paper we consider bootstrapping and subsampling. Let $X = (X_1, \dots, X_n)$ be a sample of $n$ i.i.d. instances. A resampling of $X$ using bootstrapping is defined as:
    \[
        \tilde{X} = (\tilde{X}_1, \dots, \tilde{X_n}) \qquad \tilde{X}_i \sim X \; \tu{(uniform with replacement)}
    \]
    and a resampling of $X$ using subsampling is defined as:
    \[
        \tilde{X} = (\tilde{X}_1, \dots, \tilde{X_m}) \qquad \tilde{X}_i \sim X \; \tu{(uniform without replacement)}
    \]
    where $m < n$.

\subsection{Directed Acyclic Graphs}
\label{sec:dags}

    A \textit{directed acyclic graph} (DAG) is an edge vertex graph whose vertices and edges represent variables and conditional independence statements. We use the notation $\mc G = (V, E)$ where $\mc G$ is a graph, $V$ is a vertex set, and $E$ is an edge set. In a DAG, the edge set is comprised of ordered pairs that represent directed edges\footnote{The edge $v \at w$ corresponds to the order pair $(v, w)$.} and contains no directed cycles. 

    \vskip 5mm
    
    \noindent Let $\mc G = (V, E)$ be a DAG and $v \in V$:
    \begin{align*}
        \pa{\mc G}{v} &\eq \{ w \in V \; : \; v \at w \s[7] \tu{in} \s[7] \mc G \} \\
        \ch{\mc G}{v} &\eq \{ w \in V \; : \; v \ta w \s[7] \tu{in} \s[7] \mc G \}
    \end{align*}
    are the \textit{parents} and \textit{children} of $v$.

    \vskip 5mm

    A DAG encodes graphical conditional independence between disjoint subsets $A,B,C \sube V$ which can be read off of the DAG using \textit{d-separation} and is denoted $\istate{A}{B}{C}{\mc G}$ \citep{dawid1979conditional, pearl1988probabilistic, lauritzen1996graphical}. This criterion admits the concept of a \textit{Markov equivalence class} (MEC) which---in the context of this paper---is a collection of DAGs that represent the same conditional independence relations \citep{Spirtes1993-od}.

\subsection{Causal Discovery}
\label{sec:causal_discovery}

    DAGs are connected to causality by the \textit{causal Markov} and \textit{causal faithfulness} assumptions. A DAG is \textit{causal} if it describes the true underlying causal process by placing a directed edge between a pair of variables if and only if the former directly causes the latter \citep{Spirtes1993-od}.

    \vskip 5mm

    \noindent \textit{Causal Markov}: If $\mc G = (V, E)$ is causal for a collection of random variables $X$ indexed by $V$ with probability distribution $P$, then for disjoint subsets $A,B,C \sube V$:
    \[
        \istate{A}{B}{C}{\mc G} \s[18] \Rightarrow \s[18] \istate{A}{B}{C}{P}.
    \]

    \noindent More generally, this implication is called the \textit{Markov property} and we say DAGs satisfying this property \textit{contains} the distribution.

    \vskip 5mm

    \noindent \textit{Causal faithfulness}: If $\mc G = (V, E)$ is causal for a collection of random variables $X$ indexed by $V$ with probability distribution $P$, then for disjoint subsets $A,B,C \sube V$:
    \[
        \istate{A}{B}{C}{P} \s[18] \Rightarrow \s[18] \istate{A}{B}{C}{\mc G}.
    \]

    \noindent Under the causal Markov and causal faithfulness assumptions, finding the MEC of the causal DAG is equivalent to identifying the simplest DAG(s) in terms of edge count that contains the data generating distribution. Causal discovery algorithms automate this search process.

\subsection{The Bayesian Information Criterion and Likelihood Ratio Test}

    Causal Discovery algorithms are generally characterized as score-based or constraint-based. Both approaches assess conditional independence relationships in that data---score-based methods with a goodness-of-fit score and constraint-based methods with conditional independence tests. In this paper we focus on the Bayesian Information Criterion (BIC) \citep{schwarz1978estimating, haughton1988choice} and the Likelihood Ratio Test (LRT) \citep{anderson1984introduction} for Gaussian DAG models.

    For a Gaussian model characterized with DAG $\mc G$ the BIC on $X$ is defined as:
    \begin{align*}
        \tu{BIC}(\mc G, X) &= \ell(\hat{\theta_{\mc G}}) - \frac{|E|}{2}
        \log(n) \\
        &= \sum_{v \in V} \ell(\hat{\theta_{v \mid \pa{\mc G}{v}}}) - \frac{|\pa{\mc G}{v}|}{2} \; \log(n)
    \end{align*}
    where $\hat{\theta_*}$ is the MLE. Causal discovery methods generally comparing nested models that differ by a single edge. In what follows, let $\hat{\theta_{\not\leftarrow}}$ be the MLE of the model without the edge and $\hat{\theta_{\leftarrow}}$ is the MLE of the model with the edge. When comparing two models that differ by a single edge the difference in BIC scores is computed as:
    \[
        \Delta_\tu{BIC} = \ell(\hat{\theta_{\not\leftarrow}}) - \ell(\hat{\theta_{\leftarrow}}) - \frac{1}{2} \log(n).
    \]
    Similarly, the likelihood ratio test statistic for comparing two models that differ by a single edge is computed:
    \[
        \label{eq:lrt}
        \lambda_\tu{LR} = -2 \left[ \ell(\hat{\theta_{\not\leftarrow}}) - \ell(\hat{\theta_{\leftarrow}}) \right]
    \]
    where $\lambda_\tu{LR}$ is distributed $\chi^2_1$. Noting that the MLE for a Gaussian distribution is computed as:
    \[
        \ell(\hat{\theta}) = - \frac{n}{2} \left[ \vphantom{\Big |} \log(2 \pi) + 2\log(\hat{\sigma}) - 1 \right]
    \]
    we formulate the decision criterion for choosing between the two models (selecting the model with the edge) with the BIC as:
    \[
        n \left[ \vphantom{\Big |} \log(\hat{\sigma_{\not\leftarrow}}) - \log(\hat{\sigma_{\leftarrow}}) \right] - c \, \frac{1}{2} \log(n) > 0
    \]
    where $c$ is an extra term added to discount the parameter penalty. Similarly, we formulate the decision criterion for choosing between the two models (selecting the mode with the edge) with the LRT as:
    \[
        2n \left[ \vphantom{\Big |} \log(\hat{\sigma_{\not\leftarrow}}) - \log(\hat{\sigma_{\leftarrow}}) \right] - q > 0 \qquad \tu{where} \quad \int_{0}^{q} \chi^2_1(x) \; dx = 1 - \alpha.
    \]
    Accordingly, these decision criteria have the likelihood term (up to a multiplicative constant) and are tuned with $c$ and $\alpha$ respectively.
    
\subsection{Metrics}
\label{sec:Metrics}

We use the following metrics to evaluate the learnt graphs by using an ensemble method to compare against the true data generating graph. The ensemble estimate is calculated by thresholding predicted probabilities at 0.5, i.e., considering no edge between two node pairs if the predicted probabilities are lesser than 0.5. For metrics that hint at calibration and reliability we calculate them via calculating an edge frequency table which stores the probabilites of two nodes being adjacent to each other. Then we compute the following metrics (Brier and ECE) using the edge frequency table.

\subsubsection{Brier Scores}
In order to evaluate the calibration of the resampling techniques, we treat the resampling adjacency frequencies as predictive probability, and evaluate their predictive ability using the Brier score \citep{VERIFICATIONOFFORECASTSEXPRESSEDINTERMSOFPROBABILITY}, a standard tool for evaluating the performance for binary events. The Brier score is simply the mean squared error between the predicted probability and the observed outcomes:

\vskip 2mm
    
    \begin{align}
        \text{Brier Score} = \frac{1}{n} \sum_{s=1}^{n} (f_s - o_s)^2\
    \end{align}
    \vskip 2mm
where n is the sample size, fs is the forecast probability of the event in sample s, and os is the observed outcome in sample s, with 0 meaning the event did not occur and 1 meaning the event did occur. Since this is a measure of error, larger values are worse: the best Brier score possible is 0, while the worst score possible is 1.

\subsubsection{Expected Calibration Error}
While Brier scores are commonly used to evaluate the calibration of probabilistic models, they provide an incomplete assessment of model calibration on their own. To gain a more comprehensive understanding, Expected Calibration Error (ECE) \citep{Pakdaman_Naeini_Cooper_Hauskrecht_2015} is often used in conjunction with Brier scores. ECE measures the discrepancy between predicted probabilities and observed outcomes by dividing the predictions into \textbf{\textit{K}} equally spaced bins based on their confidence scores.
Mathematically, ECE is defined as:

\begin{align}
    \text{ECE} = \sum_{i=1}^{K} P(i) \cdot |o_i-e_i|
\end{align}

where \(o_i\) is the true fraction of positive instances in bin i, \(e_i\)
is the mean of the post-calibrated probabilities for the instances in bin i, and P(i) is the empirical probability (fraction) of all instances that fall into bin i. Lower ECE values indicate better calibration, as they suggest that the predicted probabilities more accurately reflect the observed frequencies of the positive class.

\subsubsection{Precision, Recall, and F1-score}

We report Precision, Recall, and F1-score to capture model performance \citep{SOKOLOVA2009427}. These metrics offer a more detailed understanding of how well a model identifies true positive instances while minimizing false positives and false negatives. 

\textbf{Precision} measures the proportion of correctly predicted positive instances out of all instances predicted as positive. It is defined as:

    \begin{align} 
        \text{Precision} = \frac{TP}{TP + FP} 
    \end{align}

where TP denotes true positives and FP denotes false positives. High precision indicates a low rate of false positives, ensuring that the positive predictions made by the model are reliable.

\textbf{Recall}, measures the proportion of actual positive instances that were correctly identified by the model. It is defined as:

    \begin{align} 
        \text{Recall} = \frac{TP}{TP + FN}
    \end{align}

where FN denotes false negatives. High recall ensures that most of the actual positive instances are captured by the model, minimizing missed detections.

To balance the trade-off between Precision and Recall, the F1-score is often used. The F1-score is the harmonic mean of Precision and Recall, providing a single composite metric that captures both aspects of model performance:

    \begin{align} 
        \text{F1 score} = \frac{2 \times \text{Precision} \times \text{Recall}}{\text{Precision} + \text{Recall}}
    \end{align}

The F1-score ranges from 0 to 1, with higher values indicating better overall model performance. It is particularly useful in scenarios with class imbalance, where optimizing for either Precision or Recall alone may lead to suboptimal model behavior.

\section{Analysis of Resampling Techniques}
\label{sec:Methods}

    This paper reports the results of a series of simulations conducted as an extension of the work by \cite{kummerfeld2019simulationsevaluatingresamplingmethods} to enhance the understanding of applying resampling to causal discovery. 

    To simulate the data we used the DaO method \citep{andrews2024bettersimulationsvalidatingcausal} to simulate data involving two graph types.  
    
    The data were generated from a causal graphical model, followed by repeated resampling to create large collections of datasets. Each resampled dataset was analyzed using two causal discovery algorithm namely fast Greedy Equivalence Search (fGES) and Best Order Score Search (BOSS) with different hyperparameters for each, resulting in a corresponding large collection of graphs. 

    For each pair of variables, we calculated the proportion of times a specific relationship was represented by an edge (or lack of one) between them across the generated graphs. These proportions were then evaluated as predictive probabilities of whether an edge existed in the underlying data-generating model. We assessed the predictive probabilities (selecting a threshold of 0.5 for Precision, Recall, and F1 score) using several metrics, including the full Brier score, Expected Calibration Error (ECE), Precision, Recall and F1 score, comparing them to the true graphs that generated the data. 

    The process can be visualized by the picture below: 
    \begin{figure}[H]
    \centering
    \includegraphics[width=\linewidth]{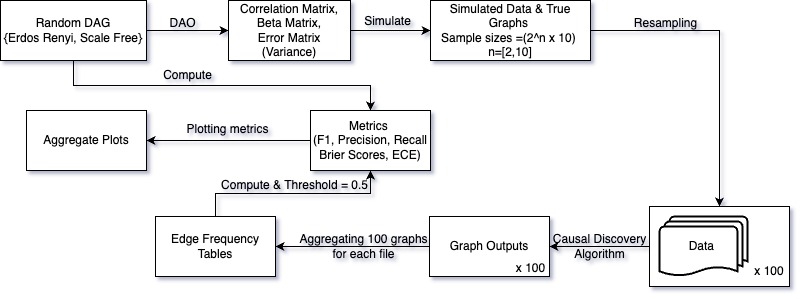}
    \caption{Simulation flow diagram}
    \label{fig:Overview diagram}
    \end{figure}

    \subsection{What, Why and How of using ESS}
    
\begin{figure}[H]
        \centering
        \includegraphics[width=0.5\linewidth]{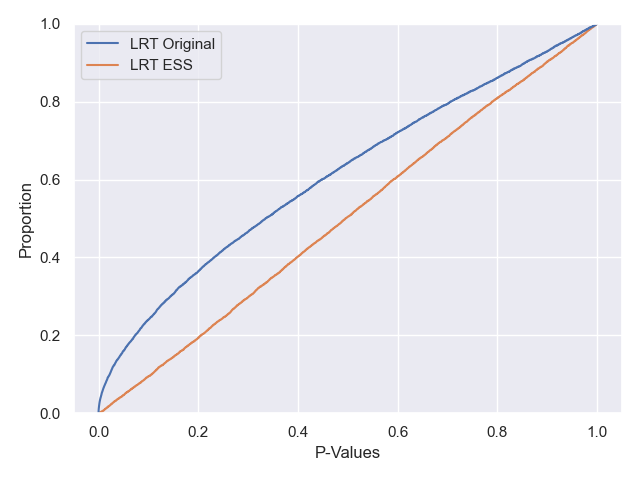}
        \caption{Non-uniformity of LRT p-values under the null hypothesis.}
        \label{fig:non-uniform_p-values}
    \end{figure}

    Let $X = (X_1, \dots, X_n)$ be a sample of $n$ i.i.d. instances and $\tilde{X}$ be a bootstrapped sample. While the number of instances in $\tilde{X}$ is still $n$, the effective sample size is smaller because of the sample contains repeated instances. To see this in practice, we consider a likelihood ratio test (LRT) on bootstrapped samples using the effective sample size and the original sample size in the LRT calculations. Figure \ref{fig:non-uniform_p-values} shows the distribution of the p-values from the likelihood ratio tests where the data were generated under the null hypothesis---two independent standard normal distributions. Notice that the p-values from the likelihood ratio tests are uniform when using the effective sample size. However, the p-values are clearly not uniform when using the original sample size.
    
    Let $w_i$ be frequency weights (number of times the $i^\tu{th}$ instance appears in $\tilde{X}$). The effective sample size is calculated as follows \citep{kish1965survey}:
    \[
        n_\tu{ess} = \frac{(\sum_{i=1}^n w_i)^2}{\sum_{i=1}^n w_i^2} = \frac{n^2}{\sum_{i=1}^n w_i^2} \approx \frac{n}{2}.
    \]
    For a Gaussian model characterized with DAG $\mc G$ we calculate the BIC on $\tilde{X}$ with $n_\tu{ess}$ as follows:
    \begin{align*}
        \tu{BIC}(\mc G, \tilde{X}) &= n_\tu{ess} \log(\hat{\theta_{\mc G}}) - c \, \frac{|E|}{2} \log(n_\tu{ess}) \\
        &\approx \frac{n}{2} \log(\hat{\theta_{\mc G}}) - c \, \frac{|E|}{2} \left[ \vphantom{\Big |} \log(n) - \log(2) \right] \\
        &= \frac{n}{2} \log(\hat{\theta_{\mc G}}) - c \, \frac{|E|}{2} \log(n) + O(1) \\
        &\propto n \log(\hat{\theta_{\mc G}}) - \tilde{c} \, \frac{|E|}{2} \log(n) + O(1)
    \end{align*}
    where $\hat{\theta_{\mc G}}$ is the MLE and $\tilde{c} = 2 c$. Accordingly, calculating the BIC on $\tilde{X}$ with $n_\tu{ess}$ is similar to using the original sample size and doubling the penalty discount term.

    \subsection{Simulation Design}
    \label{sec:simulation_design}
    To evaluate resampling techniques, we performed an in-silico experiment involving simulations. 
    
    \textbf{Random DAG}: We generated random Erd\H{o}s R\'{e}nyi and scale free DAGs with \(variables \in \{20,100\}\) and \(average\; degree \in \{2,6\}\). For stability purposes, we generate 250 true graphs, for \(variable=20\), and 50 true graphs for \(variable=100\). 
    
    \textbf{Correlation Matrix}: 
    The DAG is then used to compute a correlations matrix (and therefore a beta matrix and covariance matrix for additive error terms) using the DaO method \citep{andrews2024bettersimulationsvalidatingcausal}.
    
    \textbf{Simulated Data}: Datasets are simulated using the beta matrix and covariance matrix for additive error terms. We generated datasets for several sample sizes such that \( sample\; sizes = 2^n \cdot 10\) where \(n \in \{2, \dots, 10\}\). 

    \textbf{Resampling}: Extending \cite{kummerfeld2019simulationsevaluatingresamplingmethods}, We employed four resampling techniques: bootstrapping with the original sample size, bootstrapping with the effective sample sizes, $50\%$ subsampling, and $90\%$ subsampling. We resample each dataset 100 times\footnote{\(50 \cdot 100 = 5,000\) for 100 variables  or \(250 \cdot 100 = 25,000\) for 20 variables}. 
    
    \textbf{Causal Discovery Algorithm}: Two causal discovery algorithms were applied to each combination the simulation parameters using the Python wrapper of Tetrad \citep{ramsey2018tetrad, ramsey2023pytetradrpytetradnewpython}.
    The algorithms used in this paper are Best Order Score Search (BOSS) \citep{andrews2023fast} and Fast Greedy Equivalence Search (fGES) \citep{chickering2002optimal, ramsey2017million}. These methods search through the space of DAGs using a goodness of fit score in order to learn a model that accurately represents the data. For both algorithms we use the Bayesian Information Criterion (BIC) score \citep{schwarz1978estimating, haughton1988choice} with an added penalty discount tuning parameters. We experiment with penalty discount where \(penalty\; discount =\{1, 2\}\) 
    
    \textbf{Edge Frequency Table}: After the search completed, the graphs learnt for each iteration of the 100 resamples, were converted into edge frequency table averaging over the graphs on the basis of how frequently an edge existed in those 100 resampled graphs for each original file (50 for 100 variables and 250 for 20 variables). These frequency tables were then compared with the true graphs to compute metrics such as the Brier Score, Expected Calibration Error, F1 Score, Precision, and Recall with a threshold of 0.5.

    In our experiments, we concluded that there is no resampling technique that is superior and can be recommended in the causal discovery context. However, we did find a close and unexpected link between the resampling type and the hyperparameter, Penalty Discount.

\subsection{Simulation Results}
\label{sec:Results}

To evaluate the impact of resampling ensembles on graphical model structure learning, we conducted extensive experiments using BOSS and fGES algorithms across different network structures, variable counts, and graph densities. Our analysis focuses on multiple performance metrics including F1 score, precision, Brier score, and Expected Calibration Error (ECE).

\subsubsection{F1 Score Performance Across Resampling Techniques}

\begin{figure}[H]
    \centering
    \begin{subfigure}{0.47\textwidth}
        \includegraphics[scale=0.35]{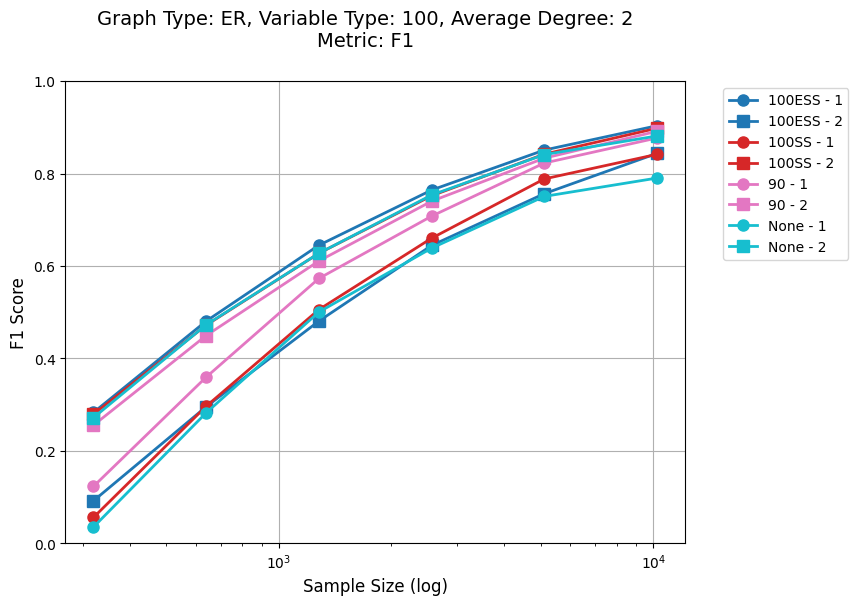}
        \caption{BOSS: 100 Variables, Average Degree 2}
        \label{fig:BOSS_F1}
    \end{subfigure}
    \hfill
    \begin{subfigure}{0.47\textwidth}
        \includegraphics[scale=0.35]{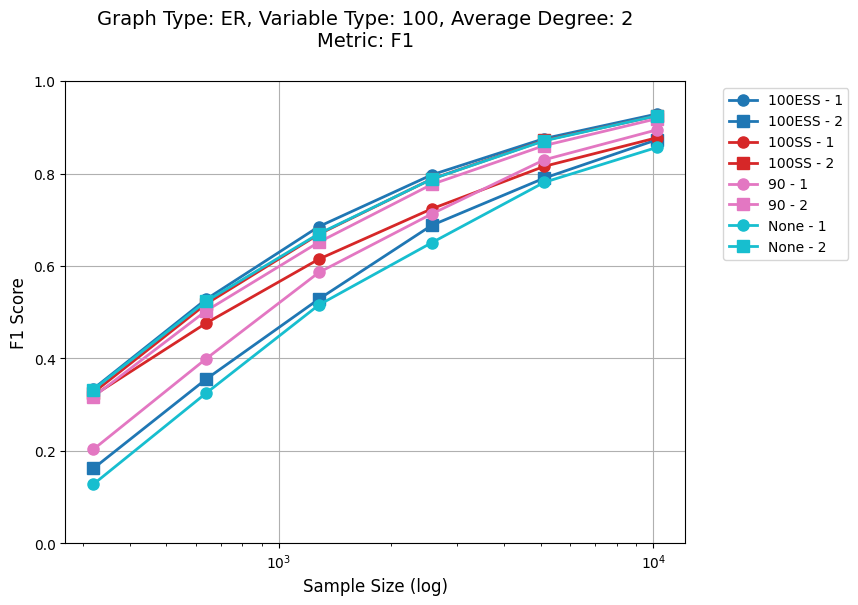}
        \caption{fGES: 100 Variables, Average Degree 2}
        \label{fig:FGES_F1}
    \end{subfigure}
    \caption{F1 scores for BOSS and fGES algorithms on ER graphs with 100 variables and average degree 2}
    \label{fig:F1_scores}
\end{figure}

Figure~\ref{fig:F1_scores} presents the F1 scores for edge recovery in Erd\H{o}s–R\'{e}nyi (ER) random graphs with 100 variables and average degree of 2. The different penalty discounts are represented by distinct markers ($\circ$ for penalty discount = 1 and $\square$ for penalty discount = 2), while resampling strategies are distinguished by color coding.

For both BOSS (Figure~\ref{fig:BOSS_F1}) and fGES (Figure~\ref{fig:FGES_F1}) algorithms, no single resampling technique consistently outperforms all others across all sample sizes and penalty discount combinations. The bootstrap with adjusted sample size (100ESS) with penalty discount 1 and the 90\% subsample approach with penalty discount 2 demonstrate particularly strong performance across most sample sizes. Notably, the bootstrap without sample size adjustment (100SS) performs relatively poorly when combined with penalty discount 1, indicating that this combination may inadequately balance model complexity and data fit.

As sample size increases to 10,240, performance differences between resampling techniques diminish but remain discernible, suggesting that resampling strategy selection remains consequential even with abundant data. These results highlight the importance of jointly optimizing both resampling approach and penalty parameter settings for effective graphical model estimation.

\subsubsection{Precision Analysis and Protection Against Erroneous Edges}

\begin{figure}[H]
    \centering
    \begin{subfigure}{0.47\textwidth}
    \includegraphics[scale=0.35]{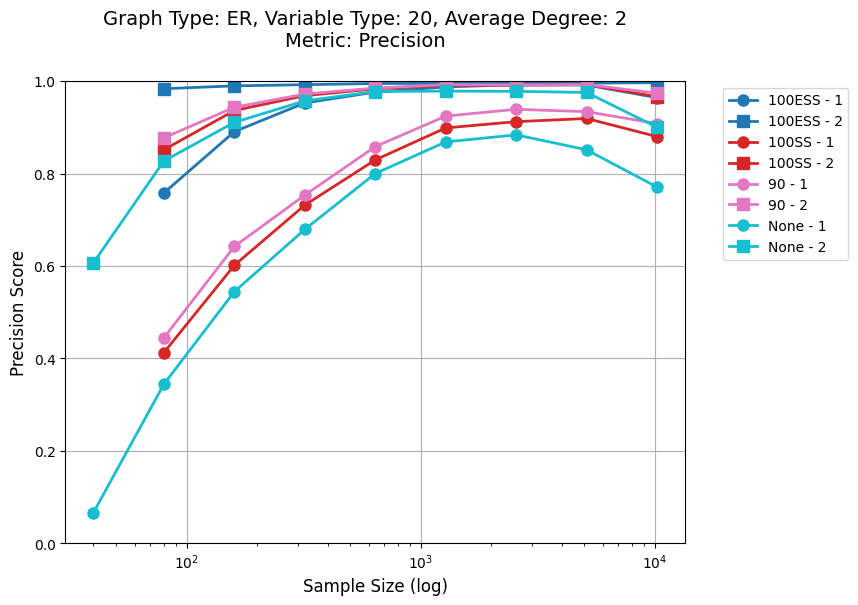}
    \caption{ER, 20 Variables, Average degree 2}
    \label{fig:BOSS_20_2_precision}
    \end{subfigure}
    \hfill
    \begin{subfigure}{0.47\textwidth}
    \includegraphics[scale=0.35]{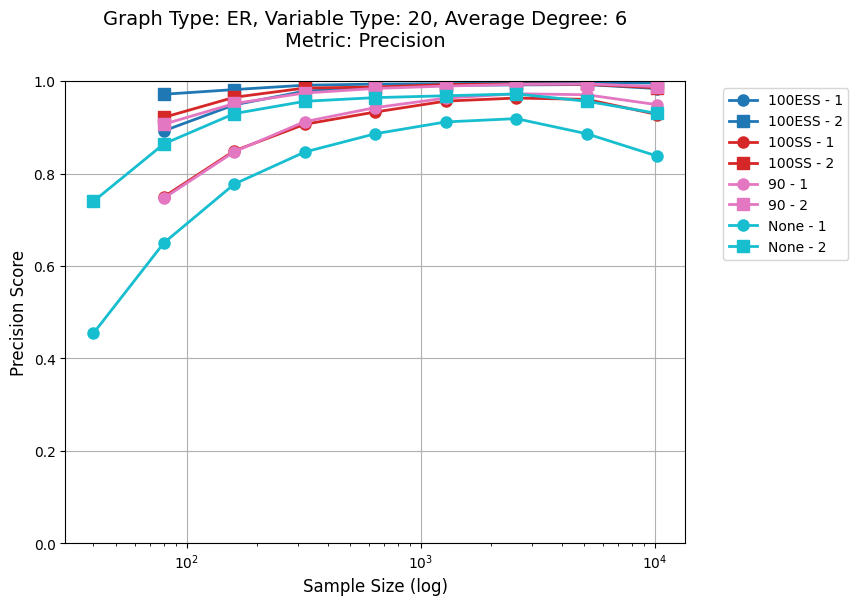}
    \caption{ER, 20 Variables, Average degree 6}
    \label{fig:BOSS_20_6_precision}
    \end{subfigure}
    \caption{Precision scores for BOSS algorithm demonstrating the protective effect of resampling against fitting noise}
    \label{fig:precision_plots}
\end{figure}

Figure~\ref{fig:precision_plots} illustrates precision scores for edge recovery in Erdős–Rényi (ER) graphs with 20 variables, comparing networks with average degrees of 2 and 6. These results demonstrate that resampling techniques provide robust protection against fitting erroneous edges, with this protective effect evident across different network densities.

For sparse networks (average degree = 2, Figure~\ref{fig:BOSS_20_2_precision}), all resampling strategies achieved precision scores above 0.8 with just 100 samples, while the non-resampled approach required approximately 1,000 samples to reach comparable performance. The bootstrap with adjusted sample size maintained precision scores near 1.0 across most sample sizes, exhibiting particularly strong resistance to fitting incorrect edges.

In denser networks (average degree = 6, Figure~\ref{fig:BOSS_20_6_precision}), the protective effect of resampling was equally evident but with overall higher baseline precision scores. Even with increased edge density, resampling strategies maintained precision scores above 0.9 at sample sizes as low as 100, while non-resampled approaches showed substantially lower precision until reaching larger sample sizes.

Critically, all methods demonstrated a slight decrease in precision at the largest sample size (10,240), potentially due to overfitting when excessive data is available. This phenomenon was more pronounced in the non-resampled strategy, providing compelling evidence that resampling protects against fitting incorrect or noisy edges even in data-rich scenarios. This protective effect is complementary to and can be enhanced by appropriate penalty discount selection, as indicated by the differential performance across penalty parameter settings within each resampling strategy.

\subsubsection{Brier Score Analysis for Calibration Assessment}

\begin{figure}[H]
    \centering
    \begin{subfigure}{0.47\textwidth}
        \includegraphics[scale=0.35]{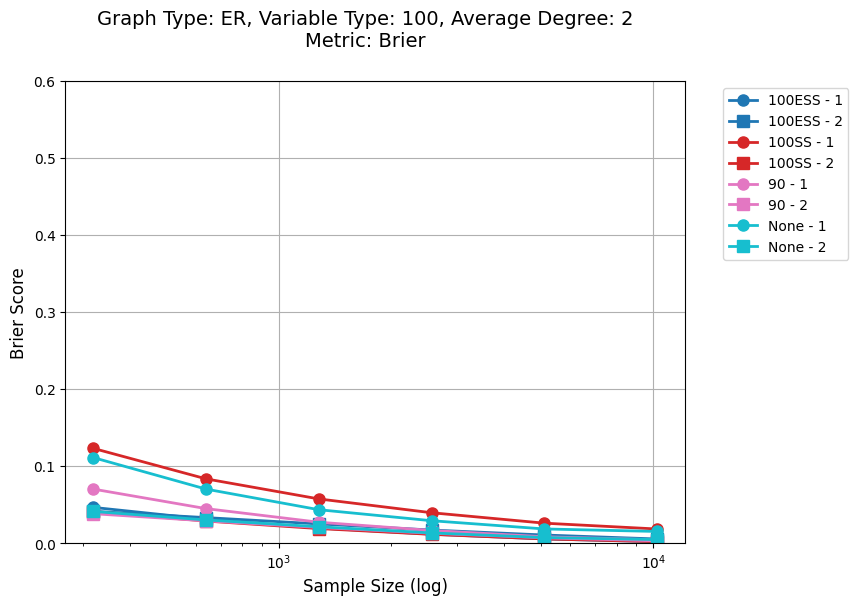}
        \caption{BOSS, 100 Variables, Average Degree 2}
        \label{fig:BOSS_Brier}
    \end{subfigure}
    \hfill
    \begin{subfigure}{0.47\textwidth}
        \includegraphics[scale=0.35]{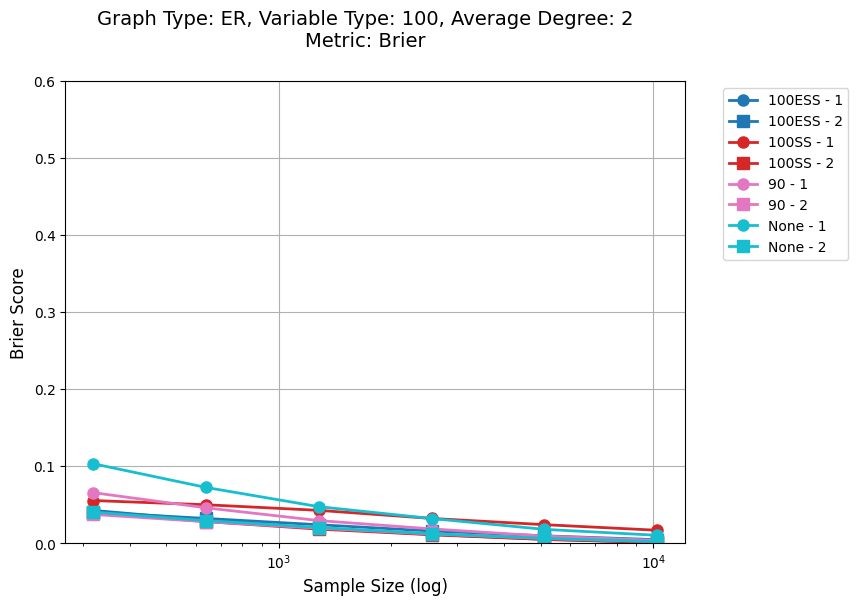}
        \caption{fGES, 100 Variables, Average Degree 2}
        \label{fig:FGES_Brier}
    \end{subfigure}
    \caption{Brier scores for BOSS and fGES algorithms demonstrating calibration quality}
    \label{fig:Brier_scores}
\end{figure}

Figure~\ref{fig:Brier_scores} displays the Brier scores for edge recovery in ER random graphs with 100 variables and average degree of 2. The Brier score, measuring probabilistic calibration reliability, shows consistent improvement (decreasing scores) as sample size increases across all resampling methods for both BOSS (Figure~\ref{fig:BOSS_Brier}) and fGES (Figure~\ref{fig:FGES_Brier}) algorithms.

At smaller sample sizes, notable differences emerge between resampling strategies and their interaction with penalty discount parameters. The bootstrap with adjusted sample size yields the lowest Brier scores (approximately 0.04-0.05), indicating superior calibration of edge probability estimates compared to other methods. In contrast, the non-resampled approach and bootstrap without sample size adjustment exhibit higher Brier scores (0.10-0.12) at low sample sizes, suggesting poorer calibration of uncertainty estimates.

As sample size increases to 10,240, all methods converge toward similarly low Brier scores (below 0.02), indicating that with sufficient data, the advantage of resampling for calibration diminishes. This convergence suggests that while resampling provides substantial benefits for uncertainty quantification in data-limited scenarios, these benefits become less pronounced with abundant data.

Notably, the bootstrap with adjusted sample size maintains better calibration across all sample sizes, which aligns with our precision analysis results. This dual advantage in both edge recovery precision and probability calibration highlights the robustness of ensemble-based resampling approaches for graphical model estimation.

\subsubsection{Expected Calibration Error Analysis}

\begin{figure}[H]
    \centering
    \begin{subfigure}{0.47\textwidth}
        \includegraphics[scale=0.35]{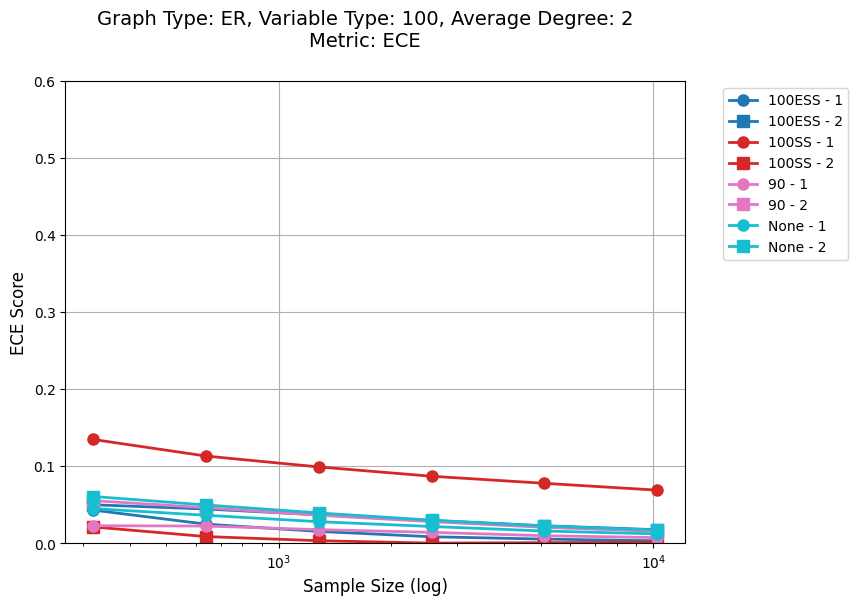}
        \caption{BOSS, 100 Variables, Average Degree 2}
        \label{fig:BOSS_ECE}
    \end{subfigure}
    \hfill
    \begin{subfigure}{0.47\textwidth}
        \includegraphics[scale=0.35]{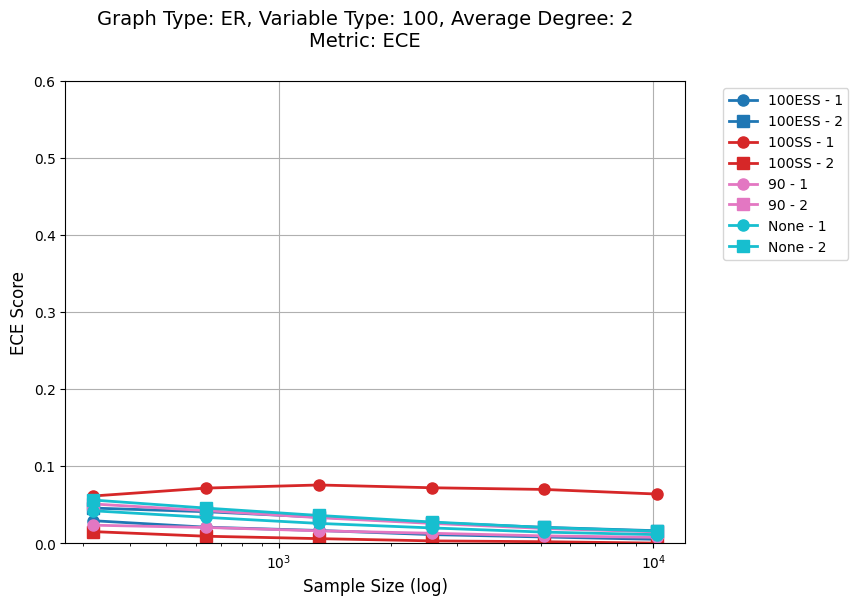}
        \caption{fGES, 100 Variables, Average Degree 2}
        \label{fig:FGES_ECE}
    \end{subfigure}
    \caption{Expected Calibration Error for BOSS and fGES algorithms}
    \label{fig:ECE_scores}
\end{figure}

Figure~\ref{fig:ECE_scores} illustrates the Expected Calibration Error (ECE) for edge recovery in both BOSS (Figure~\ref{fig:BOSS_ECE}) and fGES (Figure~\ref{fig:FGES_ECE}) with 100 variables and average degree of 2. ECE, a more fine-grained measure of calibration than Brier score, reveals distinct patterns across resampling techniques and their interaction with penalty discount parameters.

The bootstrap with adjusted sample size demonstrates consistently low ECE values (below 0.05) across all sample sizes, with particularly strong performance when paired with appropriate penalty discount parameters. This synergistic relationship between resampling approach and penalty selection highlights the importance of calibrating both aspects of the model fitting process. In contrast, the bootstrap with original sample size exhibits notably higher ECE values, particularly in the first run where ECE remains elevated (0.07-0.08) even as sample size increases to 10,240. This persistent miscalibration suggests that bootstrapping without sample size adjustment may lead to systematic bias in uncertainty estimation that cannot be fully compensated for by penalty parameter tuning alone.

The 90\% subsample approach achieves the lowest ECE scores among all tested methods, with values consistently below 0.02 across the entire range of sample sizes, demonstrating robust calibration regardless of penalty discount selection. This superior performance, particularly when combined with the higher penalty discount (squares), indicates that subsampling may provide a particularly well-calibrated probability estimates for graphical model structure learning.

Interestingly, while most methods show expected improvement (decreasing ECE) with increasing sample size, the bootstrap with original sample size exhibits relatively stable ECE values across sample sizes, suggesting that its calibration limitations persist regardless of data abundance. This finding contrasts with our Brier score results, where all methods showed convergence at large sample sizes, highlighting the value of examining multiple calibration metrics.

\section{Discussion}

The analysis across multiple performance metrics reveals several key insights about the role of resampling in graphical model structure learning. First, Our results demonstrate an interesting interchangeability effect: similar performance can be achieved by making compensatory adjustments between resampling method and penalty discount parameters. For instance, the 90\% subsample with penalty discount 2 and the bootstrap with adjusted sample size with penalty discount 1 showed comparable performance trajectories across multiple metrics, despite employing fundamentally different resampling strategies. Additionally, bootstrap without adjusting for sample size at penalty discount 1 is comparable to bootstrap with effective sample size and 90\% subsample at penalty discount 2. This suggests that practitioners have flexibility in their methodological choices and can potentially achieve equivalent results through different combinations of these parameters.

Bootstrap methods exhibit distinctly different behavior compared to other resampling techniques. While the bootstrap with adjusted sample size demonstrates good calibration and precision, the bootstrap with original sample size shows persistent miscalibration that cannot be fully compensated for by parameter tuning alone. This unique pattern implies that bootstrap requires special consideration when deployed in graphical model estimation.

Our findings suggest different best practices depending on specific algorithmic choices and objectives. When using BOSS, the bootstrap with adjusted sample size paired with lower penalty discount values provides optimal precision. In contrast, fGES achieves superior calibration using the 90\% subsample approach with higher penalty discount parameters. These algorithm-specific findings provide valuable guidance for causal discovery practitioners.

Resampling consistently provides protection against fitting erroneous edges, as evidenced by higher precision scores across different network densities and sample sizes. This protective effect is particularly valuable in high-dimensional settings and persists even in data-rich scenarios.

Notably, the effectiveness of resampling techniques remained consistent across different network structures (Erd\H{o}s R\'{e}nyi or Scale Free) and densities for larger networks ($\geq 100$ variables). The protective effects against erroneous edge inclusion were observed in both sparse and dense networks, suggesting that graph structure has minimal impact on the relative performance of resampling methods. This consistency simplifies the search space for best practices, as recommendations can be made independently of the underlying network structure.

Resampling techniques such as the 90\% subsample and the bootstrap with adjusted sample sizes not only improve precision but also provide better calibration, as demonstrated by lower Brier scores and ECE values, particularly at smaller sample sizes. The 90\% subsample approach and bootstrap with adjusted sample size consistently exhibited superior calibration across sample sizes, underscoring the importance of appropriate sample size consideration during resampling for reliable uncertainty quantification.

These findings collectively suggest that careful selection of resampling strategy, coupled with appropriate parameter tuning, provides a robust framework for reliable graphical model estimation. The complementary protective effects of 90\% subsample and the bootstrap with adjusted sample sizes against erroneous edge inclusion and miscalibration offer substantial benefits for structure learning applications, particularly in high-dimensional settings with limited sample sizes.

\section{Future Work}

Several promising directions for future research emerge from this study. First, developing automated methods for jointly optimizing resampling techniques and penalty parameters represents a significant opportunity. While our results demonstrate that different combinations can yield comparable performance, the search for optimal parameter settings remains largely manual and heuristic. Future work could explore how resampling techniques themselves might be leveraged to determine optimal tuning parameters through methods that predict optimal parameter combinations based on data characteristics.

Additionally, investigating the theoretical properties of combined resampling and tuning parameter strategies could provide deeper insights into their complementary nature. Understanding the mathematical relationship between these components could lead to formal guarantees on performance improvements and potentially reveal optimal combinations under specific conditions.

Finally, extending our analysis to non-Gaussian and discrete data models would broaden the applicability of our findings. Different distributional assumptions might interact differently with resampling strategies, potentially revealing additional nuances in best practices for structure learning in these contexts.

\bibliographystyle{apalike}
\bibliography{refs}

\end{document}